\documentclass[
  ,draft            % use draft while you are working on the paper
  ,numberedheadings % uncomment this option for numbered sections
  ]
  {aipproc}

\layoutstyle{6x9}

\begin{document}

\title{Colliding wind binaries and $\gamma$-ray binaries\,: relativistic version of the RAMSES code}


\author{Astrid Lamberts}{
  address={UJF-Grenoble 1 / CNRS-INSU, Institut de Plan\'{e}tologie et d\textquoteright Astrophysique de Grenoble (IPAG)\\ UMR 5274, Grenoble, F-38041, France}
}

\author{Guillaume Dubus}{
  address={UJF-Grenoble 1 / CNRS-INSU, Institut de Plan\'{e}tologie et d\textquoteright Astrophysique de Grenoble (IPAG)\\ UMR 5274, Grenoble, F-38041, France}
}

\author{S\'ebastien Fromang}{
  address={Laboratoire AIM /CEA/DSM-CNRS-Universit\'e Paris Diderot, IRFU/Service d\textquoteright Astrophysique\\ CEA-Saclay F-91191 Gif-sur-Yvette, France}
}

\author{Geoffroy Lesur}{
  address={UJF-Grenoble 1 / CNRS-INSU, Institut de Plan\'{e}tologie et d\textquoteright Astrophysique de Grenoble (IPAG)\\ UMR 5274, Grenoble, F-38041, France}
}

\begin{abstract}
%Collding stellar winds and $\gamma$-ray binaries share a common structure. We investigate the stability of colliding wind binaries using numerical simulations. 
$\gamma$-ray binaries are colliding wind binaries (CWB) composed of a massive star a  non-accreting pulsar with a highly relativistic wind. Particle acceleration at the shocks results in emission going from extended radio emission to the $\gamma$-ray band. The interaction region is expected to show common features with stellar  CWB. Performing numerical simulations with the hydrodynamical code RAMSES, we focus on their structure and stability and find that the Kelvin-Helmholtz instability (KHI) can lead to important mixing between the winds and destroy the large scale spiral structure.  To investigate the impact of the relativistic nature of the pulsar wind, we extend RAMSES to relativistic hydrodynamics (RHD). Preliminary simulations of the interaction between a pulsar wind and a stellar wind show important similarities with stellar colliding winds with small relativistic corrections.

%We investigate the hydrodynamics of the interaction between two winds from companion stars. The collision results in a double shock region which is expected to turn into a spiral due to orbital motion.  This structure is expected to be common to binaries composed of massive stars and gamma-ray binaries harbouring a young pulsar. High resolution numerical simulations are required to model it and the different instabilities which develop. We use hydrodynamical code RAMSES with Adapted Mesh Refinement (AMR) to perform large scale simulations of colliding stellar winds. We find that in some configurations the Kelvin-Helmholtz instability disrupts the spiral structure which has observational consequences.  We extended RAMSES to model multidimensional relativistic flows with AMR. 
%After a brief overview of our numerical implementation, I will present preliminary simulation that contrast the structure of the flow in gamma-ray and stellar binaries. This new relativistic code is suited for the study of different high energy phenomena such as relativistic jets, pulsar wind nebulae or  gamma-ray bursts. 

\end{abstract}

\maketitle

%%%%%%%%%%%%%%%%%%%%%%%%%%%%%%%%%%%%%%%%%%%%
%% MAINMATTER
%%%%%%%%%%%%%%%%%%%%%%%%%%%%%%%%%%%%%%%%%%%%
\section{Introduction}

$\gamma$-ray binaries share a common structure with colliding wind binaries composed of two massive stars.  In the latter,  the interaction of the winds creates a double shock structure which geometry depends on the momentum flux ratio of the winds. Extensive numerical studies have studied the different  instabilities that arise in the colliding wind region \citep{2009MNRAS.396.1743P,PaperI}. At larger scale, a spiral structure is expected but its exact geometry is still being studied \citep{2011A&A...527A...3V}.
Relativistic simulations of $\gamma$-ray binaries reveal a structure similar to colliding stellar winds \citep{2012MNRAS.419.3426B} and the development of instabilities \citep{2012A&A...544A..59B}. Still, the exact impact of the relativistic nature of the pulsar wind has never been established. The aim of our work is to highlight the similarities and differences between $\gamma$-ray binaries and colliding stellar winds using numerical simulations.

\section{Stellar colliding wind binaries}

RAMSES is a second order Godunov method, that solves the equations of hydrodynamics. We use adaptive Mesh Refinement (AMR) to increase resolution and properly model the instabilities while simulating  the large scale structure at reasonable computational cost. The winds are generated following the method by \citet{Lemaster:2007sl}. We include a passive scalar to  determine mixing between the winds.

\begin{figure}[h]
  \centering
  \includegraphics[width = .25\textwidth]{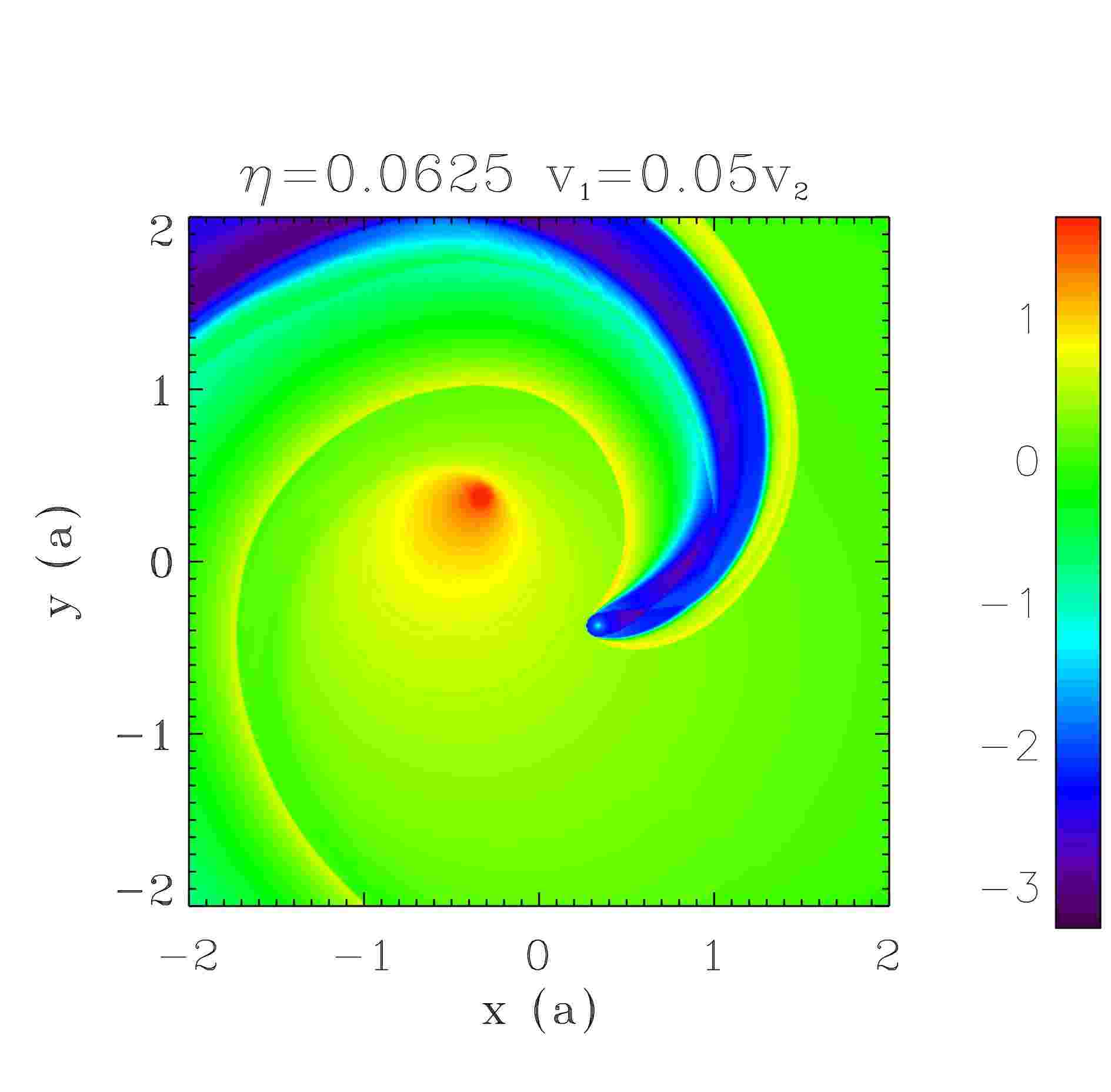}
  \includegraphics[width = .25\textwidth]{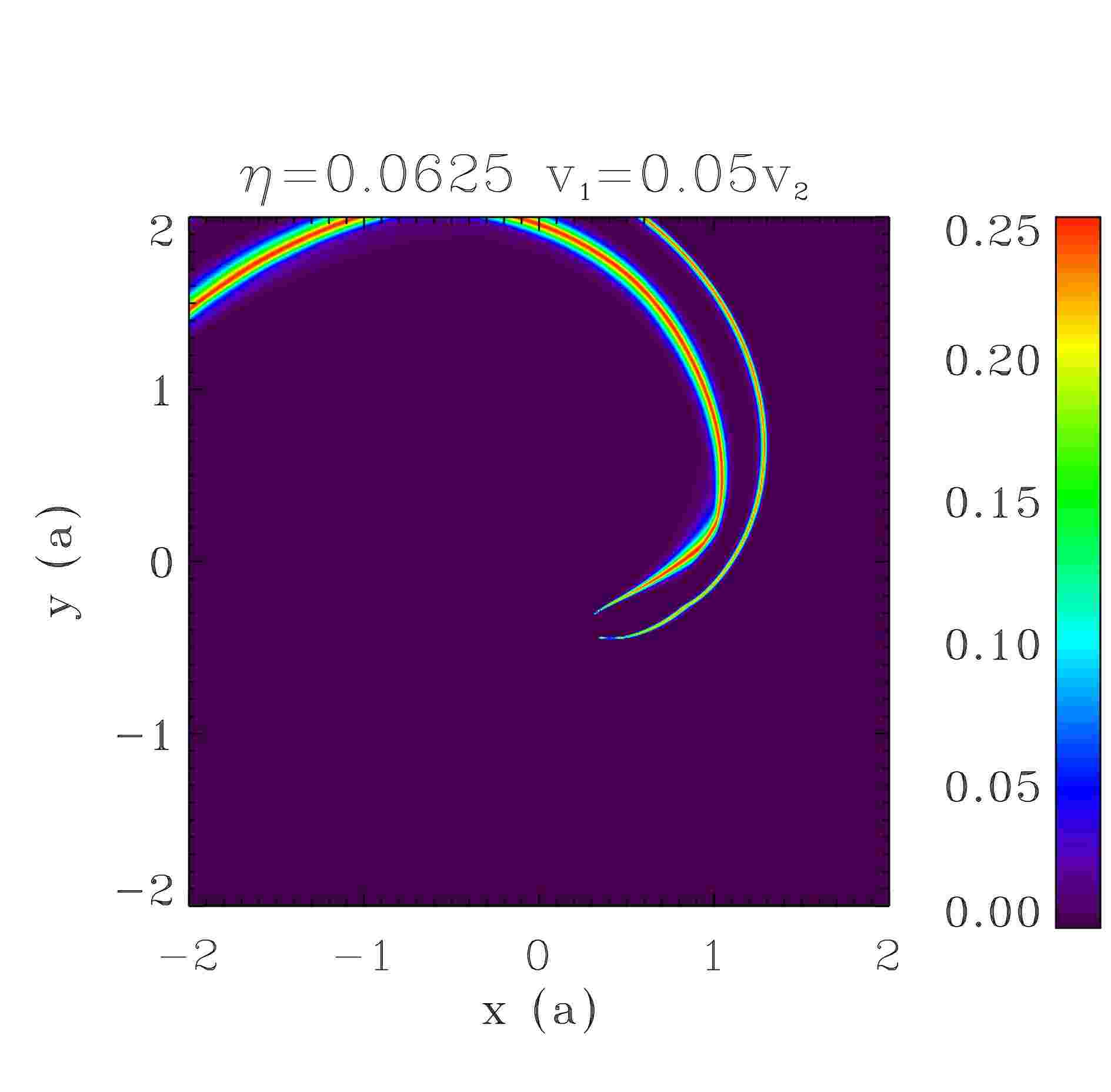}
  \includegraphics[width = .25\textwidth]{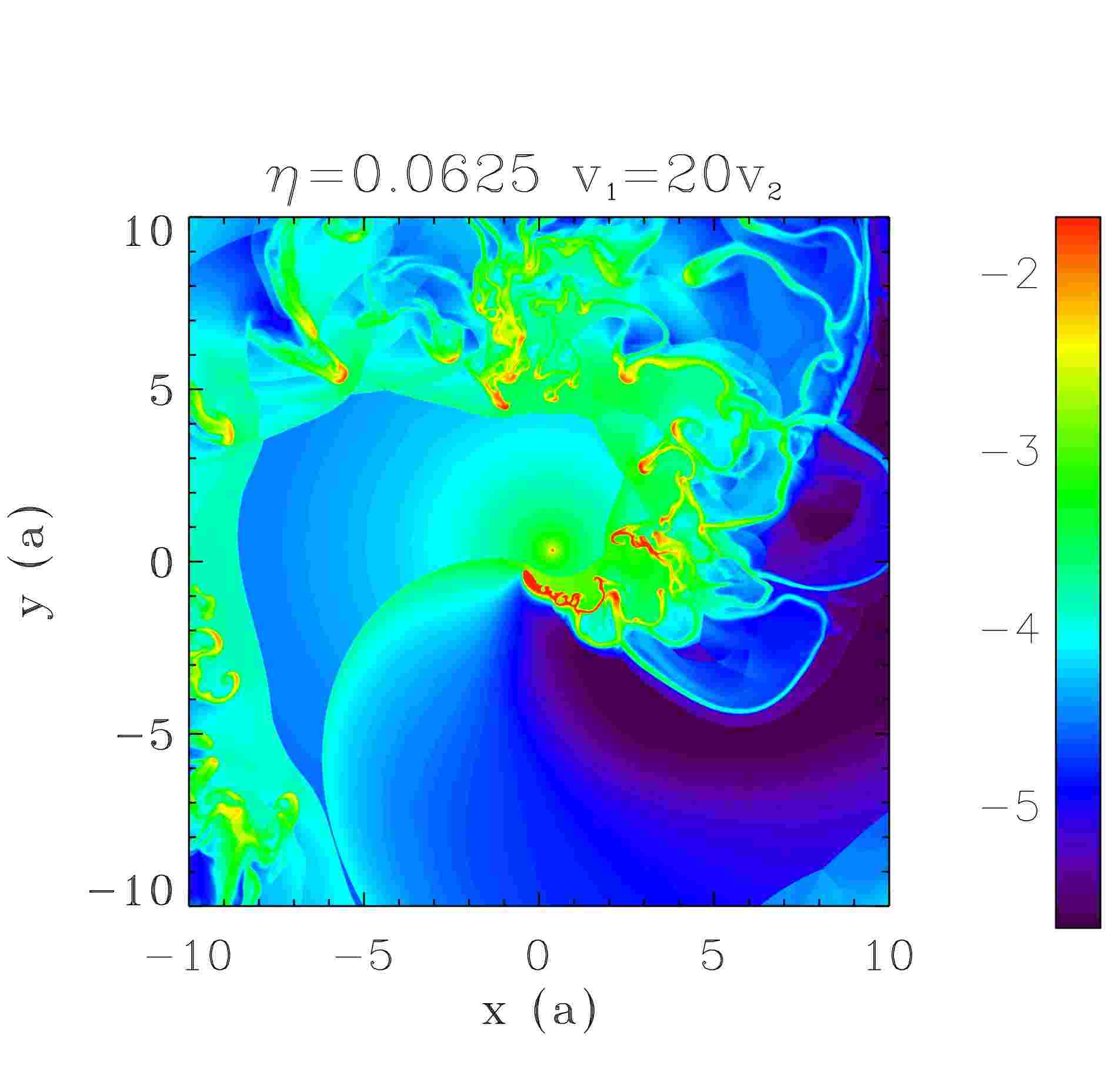}
  \includegraphics[width = .25\textwidth]{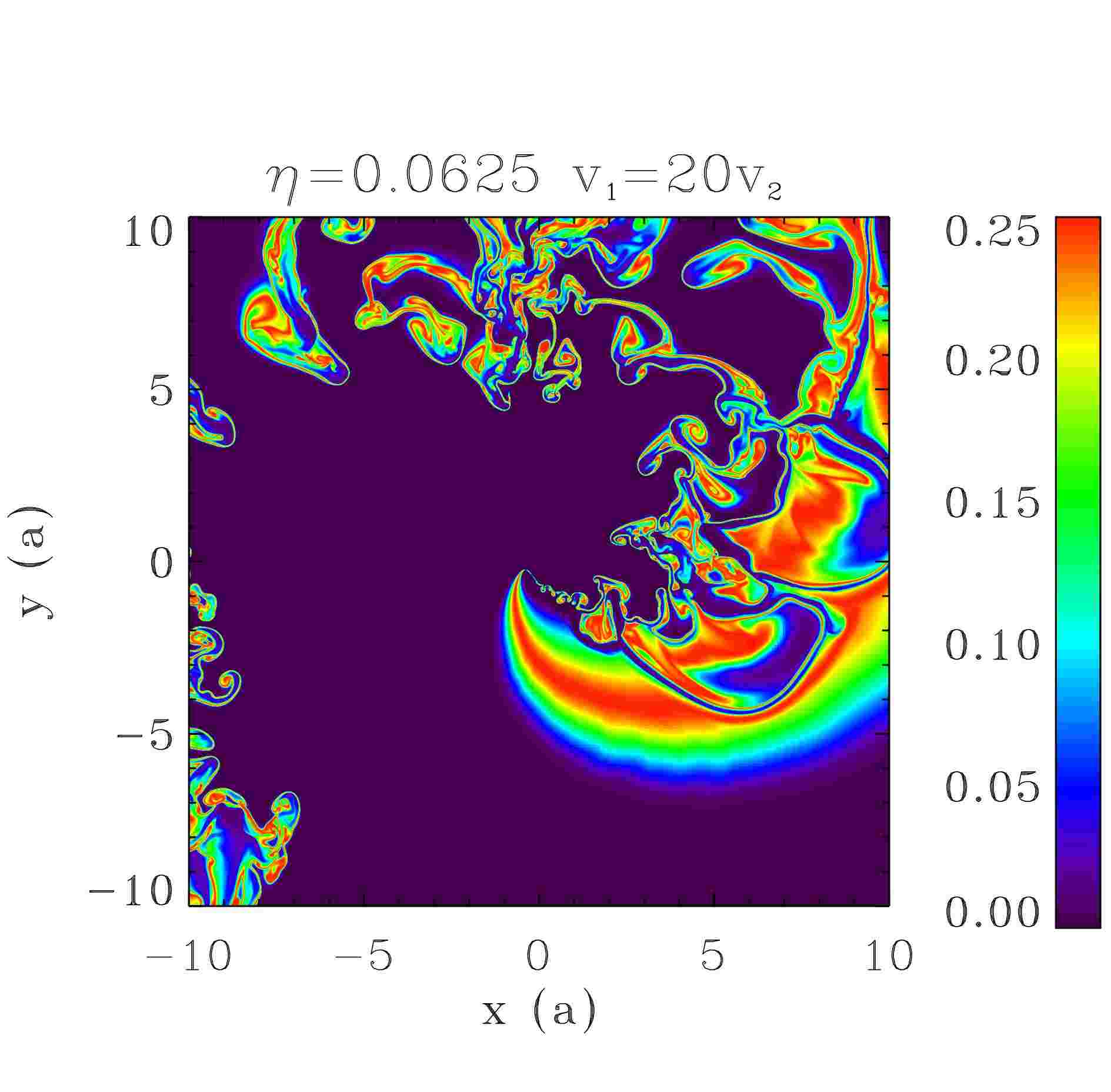}
  \caption{ Density  and mixing  for 2D simulations. In the first panels $v_1=20v_2$, in the last panels $v_1=1/20=0.05v_2$. The length scale is the binary separation.}
  \label{fig:small_scale16}
\end{figure}
Fig.\,\ref{fig:small_scale16} shows density and mixing for winds with moment flux ratio $\dot{M_2} v_2/\dot{M}_1v_1=0.0625$ and inverted velocity ratios.  The spiral structure has different properties according to the velocity of the  dominant wind $v_1$.  Using additional simulations, we determine that the step of the spiral is mostly set by $v_1\times P$ where $P$ is the orbital period of the system.  Still, the weaker wind can account for a non-negligible deviation to  this value. There is a distinction between both spiral arms, clearly visible when $v_1=20v_2$. The spiral arm propagating into the faster, lower density wind expands, while the arm propagating into the denser and slower wind gets compressed. Mixing is more important in the wider arm \citep{PaperII}. When  $v_1/v_2=20$, the spiral structure is destroyed due to the KHI while the structure is stable when $v_1/v_2=0.05$ owing to the important density gradient between the winds. 

These simulations indicate the large scale structure of CWB is strongly dependant on the wind properties and that the KHI can account for the destruction of the large scale structure in some cases.  It also leads to important mixing between the winds. In $\gamma$-ray binaries, it could enhance cooling of non-thermal particles and affect the large scale synchrotron emission \citep{2011A&A...533L...7M}.   To determine the impact of the relativistic pulsar wind on the structure and stability of the colliding wind region, we have extended RAMSES to RHD.

\section{Relativistic hydrodynamics with RAMSES}

The equations of RHD can be written as a system of conservation equations ($c\equiv 1$)\, :
\begin{equation}\label{eq:RHD}\nonumber
%\[
\begin{array}{ccc}
%\begin{aligned}
\frac{\partial{D}}{\partial{t}}+\frac{\partial{(Dv_j)}}{\partial{x_j}}&=&0\\% \label{eq:rel1}
\frac{\partial{M_i}}{\partial{t}}+\frac{\partial{(M_iv_j+P\delta_{ij})}}{\partial{x_j}}&=&0\\%\\ \label{eq:rel2}
\frac{\partial{E}}{\partial{t}}+\frac{\partial{(E+P)v_j}}{\partial{x_j}}&=&0 %\label{eq:rel3}
\end{array}
\quad
\textrm{with}
\quad
% \mathbf{U}=
\left(
 \begin{array}{c} 
D \\ 
M_i\\
E 
\end{array}
\right)
=
\left(
\begin{array}{c}
\Gamma \rho \\ %\nonumber 
\Gamma^2 \rho h v_i\\ %\nonumber 
\Gamma^2\rho h -P% \nonumber 
\end{array}
\right)
%\]
\end{equation}
where D is the mass density, $\mathbf{M}$ the momentum density and E the energy density in the frame of the laboratory.  The subscripts $i,j$ stand for the dimensions, $\delta_{i,j}$ is the Kronecker symbol. $h$ is the specific enthalpy, $\rho$ is the proper mass density, $v_i$ is the fluid three-velocity, $P$ is the gas pressure and $\gamma$ the adiabatic index. The Lorentz factor $\Gamma$ is given by $\Gamma=(\sqrt{1-v^2})^{-1/2}$. These equations have a similar structure to the equations of hydrodynamics but are more complex to solve because strongly coupled to each other by the Lorentz factor and the enthalpy.  An additional numerical constraint arises from the the fact that the velocity must remain subluminal. The similarity with the equations of hydrodynamics  allows us to closely follow the algorithm implemented in RAMSES, performing  localised changes. 

\begin{figure}
  \centering
  \includegraphics[trim= 0cm 26.5cm 0cm 0cm, clip,width = .99\textwidth ]{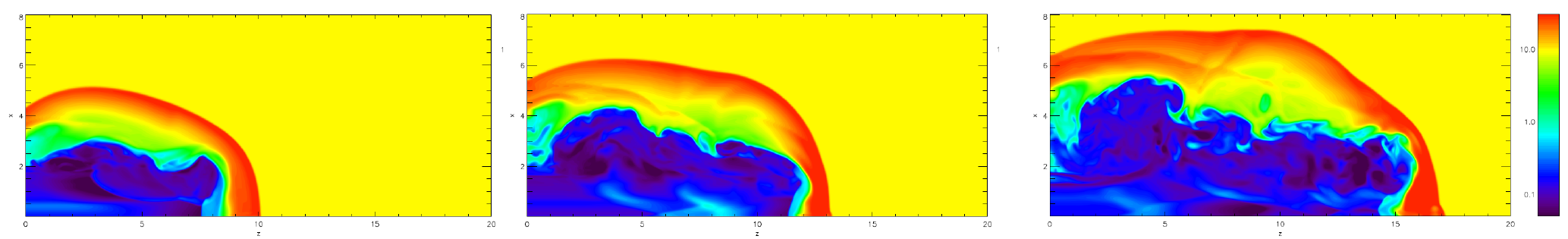}  
  \caption{ Left panel : Sod test ($t=1.8$). Right panel : Simulation of the propagation of a  3D relativistic jet ($\Gamma_{max}=7.1$). From top to bottom: density at $t=20,30,40$ in a 3D jet starting form the left boundary of the domain.}
  \label{fig:tests}
\end{figure}

We adapted the transition from the conservative variables $(D,M,E)^{T}$ to the primitive variables $(\rho,v,P)^{T}$ \citep{2007MNRAS.378.1118M}. Second-order precision is implemented in  RAMSES following a MUSCL-Hancock method and requires the determination of the Jacobian matrix of the system given by Eq.\,\ref{eq:RHD}.  The relativistic summation of velocities changes the determination of the wavespeeds and the timestep.  The implementation of RHD within the AMR structure requires adaptations when determining variables at a given refinement level $l$  by using the variables at level $l-1$ or $l+1$. Our current implementation passes the standard numerical test.  Fig.\,\ref{fig:tests} shows  the results of a 3D simulation of an axisymmetric jet following the setup by  \citet{2002A&A...390.1177D}.  This tests show satisfactory results and indicates the code is ready for scientific use. 

\section{Simulating $\gamma$-ray binaries}
\begin{figure}
  \centering
  \includegraphics[trim= .2cm .7cm 0cm 0cm, clip,width = .3\textwidth ]{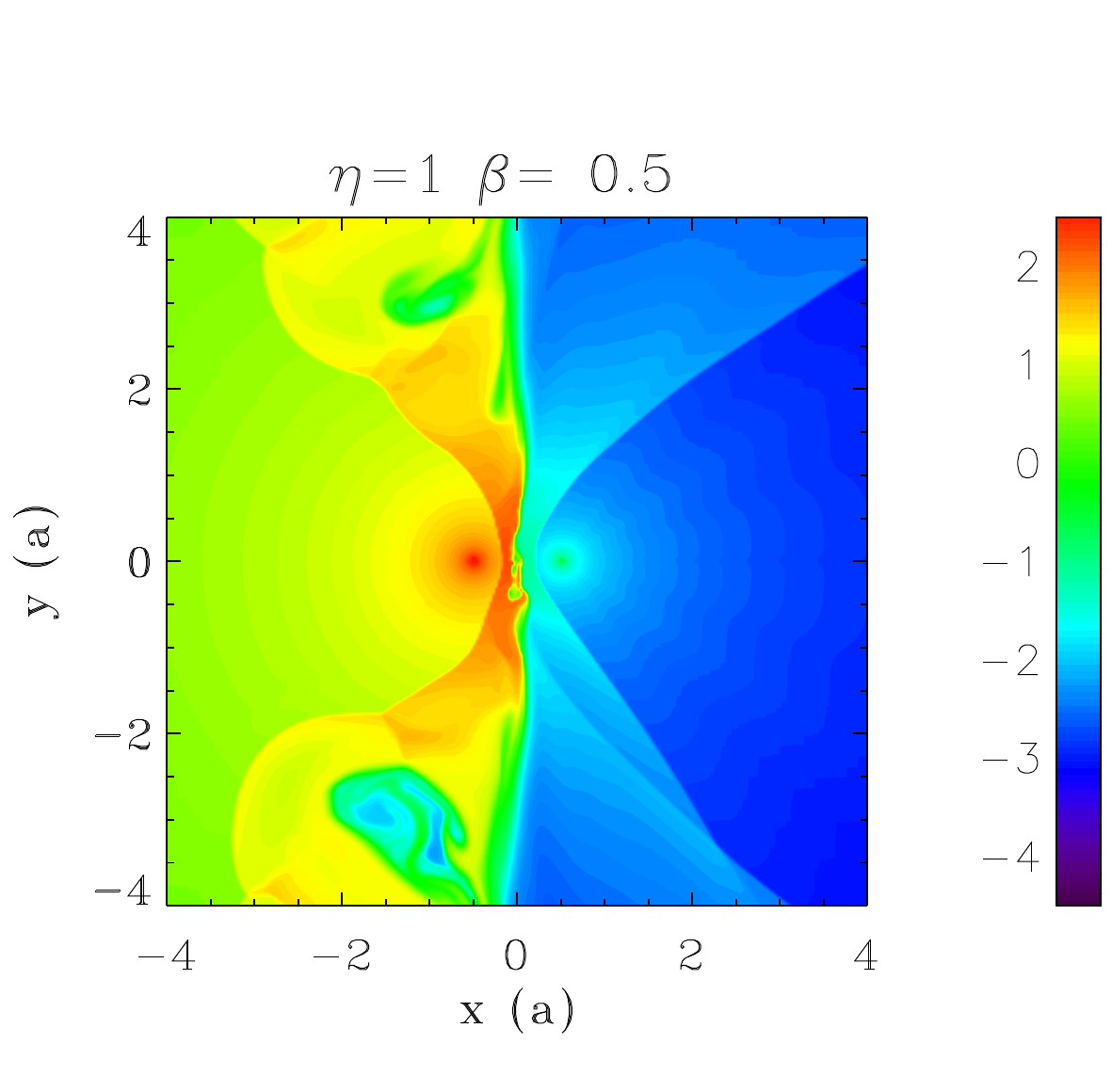}
  \includegraphics[width = .25\textwidth ]{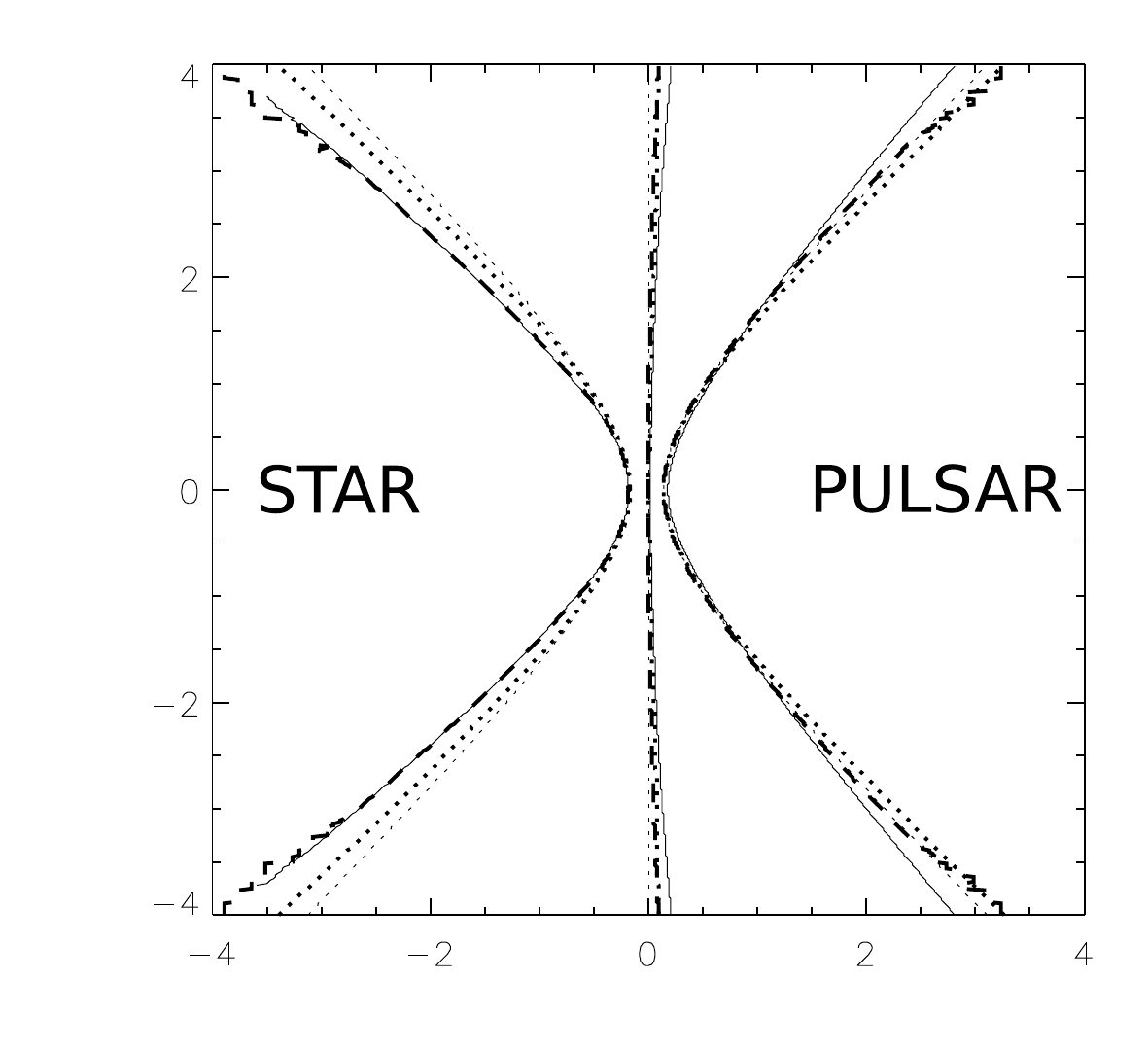}

\caption{Left panel : density map of a simulation with equal momentum flux, with $v_p=0.5$, the star is on the left, the pulsar on the right. Right panel\,: position of both shocks and the contact discontinuity, in simulations with different values for the velocity of the pulsar wind. We have $v_p$= 0.01 (thin dotted line), 0.1 (thick dotted lined), 0.5 (thick dashed line) and 0.9 (solid line). }

  \label{fig:gamma}
\end{figure}

Our aim is to model $\gamma$-ray binaries. We focus on the small scale structure of the interaction between a stellar wind and a pulsar wind. The goal is to understand the impact of relativistic effects both on the structure and stability of the interaction region.We neglect orbital motion and focus on winds with equal moment fluxes. We perform preliminary simulations with various values of the  momentum flux ratio and pulsar wind velocity. They prepare a large scale simulation to determine whether a stable structure is possible \citep{2012A&A...544A..59B}. We also want to determine the Lorentz factor downstream, as it may account for boosted emission.   Fig.\,\ref{fig:gamma} shows the density map for a simulation with the speed of the pulsar wind $v_p=0.5$. The KHI develops in a similar fashion than in the classical case. The right panel shows the positions of the discontinuities for simulations with increasing values for the pulsar wind speed. The higher  the value, the more the shocks are bent towards the star. This is a relativistic effect due to the impact of transverse velocities on the structure of shock

\section{Perpsectives}

We performed high resolution simulations of colliding wind binaries at a spatial scale never reached before. We showed that the KHI may destroy the expected large scale structure. Simulations of WR 104 match well with the observed structure and indicate cooling has to be taken into account to allow dust formation in this system. To model $\gamma$-ray binaries, we extended RAMSES to relativistic hydrodynamics. Preliminary simulations of $\gamma$-ray binaries confirm a similar structure to stellar binaries.  The relativistic extension of RAMSES allows the use of AMR and is suited for the study of gamma-ray bursts, relativistic jets or pulsar wind nebulae. It will be part of the next public release.
\begin{theacknowledgments}
AL and GD are supported by the European Community via contract ERC-StG-200911. Calculations have been performed at CEA on the DAPHPC cluster and using HPC resources from GENCI- [CINES] (Grant 2011046391)
\end{theacknowledgments}



\bibliographystyle{aipproc}   % if natbib is available

\bibliography{biblio_proceeding}

\begin{thebibliography}{10}
\expandafter\ifx\csname natexlab\endcsname\relax\def\natexlab#1{#1}\fi
\providecommand{\enquote}[1]{``#1''}
\expandafter\ifx\csname url\endcsname\relax
  \def\url#1{\texttt{#1}}\fi
\expandafter\ifx\csname urlprefix\endcsname\relax\def\urlprefix{URL }\fi
\providecommand{\eprint}[2][]{\url{#2}}

\bibitem[{Pittard}(2009)]{2009MNRAS.396.1743P}
J.~M. {Pittard}, \emph{MNRAS} \textbf{396}, 1743--1763 (2009),
  \eprint{0904.0164}.

\bibitem[{Lamberts} et~al.(2011)]{PaperI}
A.~{Lamberts}, S.~{Fromang}, and G.~{Dubus}, \emph{MNRAS} \textbf{418}, 2618
  (paper I) (2011).

\bibitem[{van Marle} et~al.(2011)]{2011A&A...527A...3V}
A.~J. {van Marle}, R.~{Keppens}, and Z.~{Meliani}, \emph{A\&A} \textbf{527},
  A3+ (2011), \eprint{1011.1734}.

\bibitem[{Bogovalov} et~al.(2012)]{2012MNRAS.419.3426B}
S.~V. {Bogovalov}, D.~{Khangulyan}, A.~V. {Koldoba}, G.~V. {Ustyugova}, and
  F.~A. {Aharonian}, \emph{MNRAS} \textbf{419}, 3426--3432 (2012),
  \eprint{1107.4831}.

\bibitem[{Bosch-Ramon} et~al.(2012)]{2012A&A...544A..59B}
V.~{Bosch-Ramon}, M.~V. {Barkov}, D.~{Khangulyan}, and M.~{Perucho},
  \emph{Astronomy \& Astrophysics} \textbf{544}, A59 (2012),
  \eprint{1203.5528}.

\bibitem[{Lemaster} et~al.(2007)]{Lemaster:2007sl}
M.~N. {Lemaster}, J.~M. {Stone}, and T.~A. {Gardiner}, \emph{Astrophysical
  Journal} \textbf{662}, 582--595 (2007), \eprint{arXiv:astro-ph/0702425}.

\bibitem[{Lamberts} et~al.(2012)]{PaperII}
A.~{Lamberts}, G.~{Dubus}, G.~{Lesur}, and S.~{Fromang}, \emph{Astronomy \&
  Astrophysics}  (2012).

\bibitem[{Mold{\'o}n} et~al.(2011)]{2011A&A...533L...7M}
J.~{Mold{\'o}n}, M.~{Rib{\'o}}, and J.~M. {Paredes}, \emph{Astronomy \&A
  strophysics} \textbf{533}, L7 (2011), \eprint{1108.0437}.

\bibitem[{Mignone} and {McKinney}(2007)]{2007MNRAS.378.1118M}
A.~{Mignone}, and J.~C. {McKinney}, \emph{Monthly Notices of the Royal
  Astronomical Society} \textbf{378}, 1118--1130 (2007), \eprint{0704.1679}.

\bibitem[{Del Zanna} and {Bucciantini}(2002)]{2002A&A...390.1177D}
L.~{Del Zanna}, and N.~{Bucciantini}, \emph{Astronomy \& Astrophysics}
  \textbf{390}, 1177--1186 (2002), \eprint{arXiv:astro-ph/0205290}.

\end{thebibliography}

\end{document}